# Large-scale Huygens' metasurfaces for holographic 3D near-eye displays

Weitao Song[1†], Xinan Liang[2†], Shiqiang Li[2], Dongdong Li[3], Ramón Paniagua-Domínguez[2], Keng Heng Lai[3], Qunying Lin[3], Yuanjin Zheng[1*], Arseniy I. Kuznetsov[2*]

1, School of Electrical and Electronic Engineering, Nanyang Technological University, 50 Nanyang Avenue, 639798 Singapore, Singapore

2, Institute of Materials Research and Engineering, A*STAR (Agency for Science, Technology and Research), 2 Fusionopolis Way, #08-03 Innovis, 138634, Singapore

3, Institute of Microelectronics, A*STAR (Agency for Science, Technology and Research), 2 Fusionopolis Way, #08-02 Innovis, 138634, Singapore

† These authors contributed equally to this work

*Corresponding author:
 Dr. Yuanjin Zheng, E-mail: yjzheng@ntu.edu.sg
 Dr. Arseniy I. Kuznetsov, E-mail: arseniy_kuznetsov@imre.a-star.edu.sg

**Abstract**

Novel display technologies aim at providing the users with increasingly immersive experiences. In this regard, it is a long-sought dream to generate three-dimensional (3D) scenes with high resolution and continuous depth, which can be overlaid with the real world. Current attempts to do so, however, fail in providing either truly 3D information, or a large viewing area and angle, strongly limiting the user immersion. Here, we report a proof-of-concept solution for this problem, and realize a compact holographic 3D near-eye display with a large exit pupil of 10mm×8.66mm. The 3D image is generated from a highly transparent Huygens' metasurface hologram with large (>$10^8$) pixel count and subwavelength pixels, fabricated via deep-ultraviolet immersion photolithography on 300 mm glass wafers. We experimentally demonstrate high quality virtual 3D scenes with ~50k active data points and continuous depth ranging from 0.5m to 2m, overlaid with the real world and easily viewed by naked eye. To do so, we introduce a new design method for holographic near-eye displays that, inherently, is able to provide both parallax and accommodation cues, fundamentally solving the vergence-accommodation conflict that exists in current commercial 3D displays.

# Introduction

Driven by the advances in mobile computing and an increasing desire from the general public to try more immersive visual experiences, near-eye 3D displays have emerged as a subject of intense study, with the particular focus on their applications in both virtual reality (VR) and augmented reality (AR). Aiming to show users the virtual scenes that imitate the real world, most of the commercial near-eye 3D display products generate 3D images with parallax cues realized by providing two sets of images, each corresponding to the separated perspective position of our two eyes. With the neural processing of the brain, the user reconstructs the scene and perceive the objects as 3D. However, a common limitation of most of these products is that they can only provide one virtual display depth, irrespective of the distance from the virtual object to the user. Therefore, when the user tries to focus on virtual objects that are supposed to be at different distances in the scene but finds a single focal distance for all of them, he or she experiences visual confusion and fatigue, particularly during prolonged usage. This phenomenon is called vergence accommodation conflict (VAC).

Several technologies, such as multiple focal planes [1], variable focal planes [2], light field [3,4] and holographic displays [5,6] have been introduced to alleviate or solve this problem. Among these, 3D holographic displays are considered to be the ultimate solution, as they intrinsically create real 3D objects, with full, continuous depth of view [7,8]. So far, most attempts to generate 3D digital holographic displays have used spatial light modulators [9,10] (such as Digital Micromirror Devices or Liquid-Crystal on Silicon) as the key enabling technology. The main limitation of these devices, however, is their large pixel size (>3.5µm), which translates into a small first order diffraction angle and a rather large higher-order diffraction noise. Consequently, in order to filter out this noise and to enlarge the limited viewing angle to a practical range, the optics for this type of display becomes bulky and cumbersome.

Recently, a new class of devices allowing wave-front manipulation with ultra-small pixels has emerged. The so-called metasurfaces are typically a single layer of metallic or high-index dielectric nanostructures, called nanoantennas, that can manipulate the phase, amplitude, and polarization of an incident light beam at subwavelength scales [11,12,13]. Compared to traditional diffractive optical elements (DOE), the subwavelength nature of these nanoantennas makes it possible for metasurface-based optical components to eliminate any higher order diffraction, thus greatly increasing the diffraction angle and the efficiency of the first order [14]. On top of that, the unique amplitude, phase, and polarization manipulation mechanisms provide more freedom to integrate multiple functions into a single element. In this regard, various optical functions had been demonstrated using metasurfaces, such as lensing [15-22], electromagnetic Brewster effect [23], beam bending [24-27], steering [28-30] and splitting [31,32], vortex beam generation [24,33], multifunction metasurface [34], holograms [31,35,36] and many others.

Thanks to the subwavelength size of its nanoantenna constituents, metasurface-based 3D holographic display holds promise of large viewing angle and high efficiency, with potentially very simple optics. However, so far this potential has not been fully exploited in previous works using metasurface holograms for display applications. Most of them, in fact, used simple 2D pictures as the image source [36-43]. The reconstruction, in these cases, was either projected on a physical screen or required a microscope to be viewed. On the other hand, those works that demonstrated full 3D reconstruction did so using a very small aperture. Thus, in these works, a microscope, or an image sensor [42,43] had to be used to capture the 3D nature of the reconstructed objects, which prevented their direct visualization by naked eye. In order to enlarge the aperture size of metasurface holograms, photolithography and nanoimprint are the two most promising fabrication techniques [12]. Indeed, using either, large area metalenses [44-46] and even 2D metasurface holograms [47] have been demonstrated.

In this work, combining large-scale metasurface with a new optical design, we experimentally demonstrate a high-resolution (11000 x 11000 pixel count) metasurface-based holographic 3D near-eye display. It has a large exit pupil of 10mm × 8.66mm and a virtual 3D scene (~50K active data points) with continuous depth ranging from 0.5m to 2m overlaid with the real world in an augmented reality fashion. Images with full depth cues can be directly viewed along with the real scene by naked eyes, which eliminates the VAC issue in 3D displays. The 3D image display device, a 4mm × 4mm size amorphous-silicon Huygens' metasurface hologram, is fabricated using 12 inch deep-UV immersion photolithography on a glass wafer, representing, to the best of our knowledge, the first demonstration of this kind of devices.

**Results and Discussion**

**Near-eye display with a metasurface hologram:** An ideal near-eye display should provide compact outlook and virtual 3D scene with a high-resolution. In AR, digital virtual information should blend in with the real world (Fig. 1a), which implies that the projected scenes from the display should have depth information. Fig. 1b and 1c illustrate the ideal AR experience. When the user focuses near, the optical axis of both eyes converges to meet on the interested position, and the crystalline lens will also be adjusted to focus on the same position. The convergence function is achieved by rotating the eye ball, and the accommodation can be adjusted by controlling the muscle around the crystalline lens. The virtual scene along with any real objects in near area will become simultaneously in-focus. At the same time, any virtual and real objects in the scene that are in the far area will be simultaneously blurred due to defocusing. The opposite would hold if the user would focus on some object far away and, therefore, ideally, the display should be able to generate virtual images with continuous depths.

When a metasurface hologram is used as the image source of a near-eye display device, some special considerations are needed for the design of the optics. Fig. 1d and 1e provide schematic illustrations of a general immersive and optical see-through near-eye display system with a metasurface hologram as the image source. When the metasurface hologram is illuminated with a coherent light source, a 3D holographic image can be generated in front of the metasurface hologram. Human eye cannot focus on the reconstructed 3D scene due to the small distance from the holographic images to the eye, thus, the images cannot be viewed clearly by naked eye. In order to allow the viewer to view the reconstructed 3D images, an eyepiece is needed to image the 3D information of the near image source far away. This eyepiece images the active area of the metasurface to a region around the eye pupil, forming a viewing window that is usually called exit pupil (or eye box). This is the area inside which the user can view the whole virtual 3D image. Fig. 1d and 1e show two general scenarios of near eye displays using a transmissive metasurface hologram as the image source for VR and AR applications, respectively. In the VR case (Fig. 1d), one eyepiece can be designed to image the reconstructed 3D object in front of the viewer while blocking the real-world path. For AR applications (Fig. 1e), a transparent optical combiner needs to be included in the eyepiece, typically a beam splitter or a waveguide. The eyepiece can image the virtual scene far away and allow the light from the real-world pass through the eyepiece simultaneously. With the help of the eyepiece, the depth of 3D objects generated directly by the metasurface hologram can be expanded to cover the full range from near to far distances and, thus, a virtual scene with continuous depths can be displayed along with the real world.

The design of the eyepiece for metasurface hologram based near-eye display is different from that of traditional near-eye displays. The display device (i.e. the metasurface hologram) and the 3D image directly reconstructed by the display device are not in the same depth. The imaging performance of both the display device and the reconstructed 3D image along with the eye box should be optimized when designing near-eye displays using metasurface holograms as the image source. Within the exit pupil, the user should completely observe the 3D scene reconstructed by the metasurface hologram. Thus, the metasurface hologram and the exit pupil should be set in the object-image conjugate relationship under the function of the eyepiece (Fig. S1). To obtain virtual objects ranging from the nearest distance to far away, real images should be reconstructed directly by the metasurface hologram, and the position of the real images should be set within the focal length of the eyepiece, which thus can be viewed as virtual images in front of the eye (Fig. S2). The relationship between the focal length of the eyepiece and other design parameters, including the generated holographic depth and the overall optical length is provided in Figs. 1f and 1g. To make the system compact, the overall optical length should be kept as small as possible. On the other hand, large generated holographic depth is better to provide the user with a good feeling of depth. Thus, when choosing the focal length of the eyepiece, the volume occupied by the reconstructed 3D image should be considered.

**Design and performance of the metasurface hologram:** Following the general design principle for holographic 3D near-eye display proposed above, a compact near-eye display setup has been designed. For the experimental proof of concept demonstration, a 4mm × 4mm metasurface hologram with pixel count of 11000 x 11000 is used as the source of the 3D image. In near-eye displays, a large exit pupil is typically preferred, as it offers better tolerance for users with different interpupillary distances and also allows swiveling within the eye sockets without vignetting or loss of image. As the human pupil size is typically around 2-8mm, the exit pupil should be larger than this size. In our case, we choose to set a minimum size as large as 10mm. To achieve this, a 2.5× magnification ratio of the hologram area is chosen in design, which can satisfy the exit pupil requirement. The desired reconstructed holographic 3D scene consists of three objects: the A*STAR logo, a dice and the Nanyang Technological University (NTU) logo, comprising more than 50k total active data points. In the scene, the object positions are set from 0.5m (for the A*STAR logo) all the way to 2m (for the NTU logo) away from the eye (0.5 diopter to 2 diopter) when viewing with the designed eyepiece. Considering the rendering image depth and optical size (see Fig. 1f and Fig. 1g), an eyepiece with a focal length of 50mm is designed. Moreover, to avoid the possible zeroth order diffraction of the hologram and/or any portion of incident light that may pass outside the hologram, both the axis of illumination and the metasurface are tilted off-axis by 30 degree with respect to the eyepiece. Figure 2a gives the complete system configuration, consisting of a laser illumination, a beam expander, a beam splitter, a concave mirror and the metasurface hologram. Here, the beam splitter and the concave mirror together act as an optical see-through eyepiece, which projects the virtual image and let light from the real world pass into the user's eye simultaneously. With this system, the exit pupil with a total size of 10mm × 8.66mm is achieved.

The phase map of the metasurface hologram is computed using the coherent ray tracing algorithm, which can provide 3D reconstruction [48] in high quality and is widely employed to produce computer-generated holograms (CGH). It treats the 3D objects as a collection of point sources with randomly distributed initial phases. A complex hologram is formed by summing up the complex field distributions of the point sources at the designed hologram plane. The phase hologram can then be obtained from the complex one by simply retaining the phase information and setting the amplitude to unity. It is important to note that, when calculating the hologram, the object-image relationship imposed by the presence of the eyepiece and the 30 degree off-axis tilt are taken into account to determine the location of these objects in the CGH calculation. The positional relationship between the objects point clouds and the hologram is shown in Fig 2b. After calculating the hologram, the information is encoded into the metasurface using the correspondence between the meta-atom geometries and the phase of the transmitted light, as detailed below.

The considered metasurface consists of a square lattice of silicon nanodisks with a constant period $P$ = 360 nm and height $h$ = 100nm and varying diameters, $D$, standing on top of a glass substrate and embedded in a PMMA polymer matrix that acts as an index-matching medium with the substrate (Fig. 2c). The size of the disks is chosen so that two resonances, namely electric dipole (ED) and magnetic dipole (MD), are excited in the disks at the operation wavelength of interest (680nm). When these two resonances are tuned, via aspect ratio of the disks [49,50], so that they overlap spectrally, it is possible to achieve full $2\pi$ phase modulation of a transmitted plane wave, while keeping near-unity transmission [26, 39, 51, 52]. The former is enabled by the doubly resonant characteristics of the structures, while the latter is enabled by the realization of the so called first Kerker's condition, leading to strong inhibition of the backward scattering, and caused by the destructive directional interference of ED and MD [52]. The phase as well as the amplitude modulation of a plane wave transmitted through the metasurface as a function of the wavelength (in the range from 500nm to 800nm) and the nanodisk diameters are shown in Fig. 2d and 2e. They were simulated using a commercial FDTD software (Lumerical FDTD). From the phase map obtained we can see that in the range from 650nm to around 700nm, $2\pi$ phase coverage while attaining high transmission can be achieved. The phase shift and transmission values at wavelength of 680nm as a function of the nanodisk diameters are shown in Fig. 2f. From there, 8 phase levels with constant phase steps of $\pi/4$ are retrieved and used to map the phase hologram, and the corresponding silicon disk diameters are 163nm, 184nm, 195nm, 203nm, 209nm, 215nm, 223nm and 250nm, respectively. Note that, since the period of the metasurface is much smaller than the designed operation wavelength, only the first order diffraction will be present for light reconstructed from the hologram fabricated based on this metasurface.

The metasurface fabrication is done on a 300-mm glass wafer using deep-UV immersion photolithography, a CMOS-compatible process that allows fabrication of large area metasurfaces, as well as their mass replication (see Figs. 3a and 3b). The detailed fabrication procedure is described in the methods section and the supplementary information (see Fig. S3). To test the hologram diffraction efficiency and transmission, a supercontinuum laser is used as a light source to illuminate the hologram. In order to cover the whole area of the hologram, the laser beam is expanded with a beam expander before it impinges onto the metasurface hologram. The transmitted holographic image is collected by a condenser lens with high numerical aperture and measured in the wavelength range from 500 nm to 800 nm in steps of 2nm. Optical efficiencies of the real image, the virtual image and the zeroth order diffraction are shown in Fig. 3c. A highest efficiency of above 80% is achieved at the wavelength of 728nm, accompanied with a very strong suppression of the zeroth order (<10%) and a slight contribution from the virtual image. Note that this optimum wavelength is redshifted with respect to the designed wavelength. We believe that this deviation may be due to a variation in the optical parameters of the material, caused by the deposition conditions, as well as size deviations from the design (see

Fig. S3). While it can be optimized in the future, in the current sample it has an influence on the total transmission (Fig. 3d), leading to its drop to around 20% at the wavelength of the highest diffraction efficiency.

**Performance of the near-eye displays using the metasurface hologram:** A prototype of a holographic 3D near-eye display based on the designed parameters mentioned above was set up using the designed metasurface hologram as the image source (see Fig. S4). Since, after the spectral shift observed in the experiment, the wavelength for the best diffraction efficiency cannot be easily observed by a human eye or captured by a normal camera (which filters out light at wavelengths above 700nm), a monochromic 16-bit SCMOS camera (PCO. edge 5.5, PCO AG) with a wavelength coverage from 300nm to 1100nm was used to capture the display performance at the best-efficiency wavelength. In this case, we first characterize it in a virtual reality setting, which can be obtained by simply blocking the real-world optical path. In the setup, the focal position of the camera can be controlled to focus at different depths. The performance of the holographic display at the wavelength of 728nm is shown in Fig. 3e and 3f. Due to the depth information of the generated objects, when the camera focuses on the near region, the A*STAR logo and the letters inside it are clear, while NTU logo is blurred owing to defocusing. Conversely, when focusing far away, the NTU logo becomes sharper and the A*STAR logo blurs. This indicates real 3D image reconstruction from our metasurface hologram with full depth cues. One should note, however, the presence of a slight background noise along with the reconstructed virtual images. We believe it to be mainly generated by the coherent speckle noise, which, as discussed in previous works [53], may arise from the resonant scattering on the metasurface.

From the efficiency curve in Fig. 3c, another peak in the visible range can be observed, at the wavelength of 544nm. Note, however, that the actual total transmission for this non-optimized case is actually quite low. Nevertheless, it is interesting to realize that, from the analysis of the phase shift and transmission values at this wavelength (Figs. 2d and 2e), it turns out that the previous, 8-phase-level metasurface designed for 680nm, actually acts as a binary hologram at 544nm. Indeed, for binary holograms, the diffraction efficiency of real and virtual images should be close to each other, as it is the case here (see the measured data in Fig. 3c). Since, despite the total low transmission at this wavelength, the efficiency values are reasonable, and they work at a wavelength for which a human eye is particularly sensitive, we decide to mimic a human eye by placing a simple visible light camera at the position of the exit pupil. As the wavelength of the light source is changed to 544nm, the position of the eyepiece has also to be adjusted, due to the depth difference for different wavelengths. Fig. 4a and Fig. 4b show the virtual information displayed by the system, as captured when the camera is focused at 0.5m (2 diopter) and 2m (0.5 diopter) depth, respectively. The generated 3D images are very clear with depth cues, and they can be directly observed by human eyes. To further verify the proposed near-eye display method, we move to an augmented reality configuration, in which the real-world path is unobstructed. In this situation, one toy plane model and one toy car model act as reference objects to measure the integration with the virtual images generated by the prototype. We locate them, respectively, at the distance of 50cm and 2m away from the exit pupil of our near-eye display. Fig. 4c and Fig. 4d give the augmented information displayed, as captured when the camera is focused at 0.5m (2 diopter) and 2m (0.5 diopter) depth, respectively (see also Supplementary Movie 1 to visualize the continuous change of focus at the depths ranging from 0.2m to 3m). As can be seen in Fig. 4c, when the camera focuses near, both the A*STAR logo and the toy plane model become simultaneously in focus (see lower panels for sharpness details), while both farther virtual and real objects (the NTU logo and the toy car model) become blurred. The situation is exactly reversed (NTU and car are in focus, A*STAR logo and plane are blurred) when the camera focuses far away, ultimately implying that the present solution can indeed solve the vergence - accommodation conflict issue. For completeness, generated images at different wavelengths ranging from 530nm to 600nm have been captured using exactly the same setup and different source wavelengths (see Supplementary Note 2 and Fig. S5).

## Conclusions

In summary, we have proposed and developed a proof-of-concept solution for holographic 3D near-eye display using a large-scale transmissive metasurface hologram as the image source. This metasurface-based device may rise as an alternative to current solutions, opening new avenues for research, as they enable the realization of continuous depths, which can be viewed directly with naked eye. The proposed approach can solve the visual confusion and fatigue problem in consumer products, ultimately making long-term usage of mobile AR/VR applications possible. This might help to push the adoption of near-eye displays, called to replace smart phones in the consumer electronics market, and even go beyond them, opening potential applications in education, medicine, or engineering, to mention some.

## Methods

**Metasuface hologram fabrication:** A 100 nm thick amorphous silicon film is deposited on a 12-inch glass wafer with plasma enhanced chemical vapor deposition (PECVD). The refractive index and absorption coefficient of the film are measured by ellipsometry and further used in the simulations. Following that, the pattern of the designed dielectric metasurface hologram is generated via deep ultra-violet (193 nm ArF) immersion lithography. After resist development and critical dimension inspection, the wafer is then diced into small pieces with one copy of hologram on each piece. In this way, we can fine tune the processing parameters and characterize them individually. The pattern is then transferred from the photoresist to the amorphous silicon layer by reactive ion etching with fluorine chemistry, which is a pseudo Bosch process using SF6 and C4F8 gases[54], followed by resist removal via oxygen plasma and wet cleaning process. Finally, samples are planarized by spin-coating of PMMA A5 resist (MicroChem Inc.; speed: 1000 rpm; time 60 seconds).

**Characterization and measurements:** Scanning electron microscopes (Hitachi SU8220) is used to capture the size and morphology of the nanostructures. To measure the diffractive efficiency of the metasurface hologram, a supercontinuum source (SuperK EXTREME, NKT Photonics) and multi-wavelength filter (SuperK SELECT, NKT Photonics) are used as the light source. An optical power meter (Thorlabs PM320E Model Console + S120C Detector) is utilized to measure the power of the incident light, the real image, the virtual image and the zeroth order. An infrared camera (PCOedge 5.5, PCO AG) and a single lens reflex camera (Canon EOS 80D with a standard lens of EF-S 18-135mm) are used at the position of the exit pupil to capture the display performance with the continuous depth. These two cameras are employed for infrared and visible light, respectively.

## Supplementary Materials

Fig. S1. Schematic and specific parameters of a near-eye display using a metasurface hologram.
Fig. S2. Schematic of the reconstruction depths of the holographic 3D scene provided by the metasurface.

Fig. S3. The fabrication error between the target and actual parameters.
Fig. S4. Optical measurement setup.
Fig. S5. Display performance at different wavelengths.
Supplementary Note 1. Design considerations and analysis for near-eye display set up using a metasurface hologram.
Supplementary Note 2. Characterization of the see-through display at different wavelengths.
Supplementary Movie 1. Display performance with the continuous change of focus from near to far distances.

## Acknowledgments

**Funding:** This work was supported by A*STAR RIE2020 AME Programmatic Funding (A18A7b0058) and National Research Foundation of Singapore under Grant No. NRF-NRFI2017-01.

**Author contributions:** W.S. and X.L. contributed equally to this work. W.S., X. L.,Y.Z. and A. I. K. conceived the idea and designed the experiments; S. L., D. L., R. P. D, K. H. L., J. Q. L. performed the design, optimization and fabrication of the metasurface. W. S., X. L., R. P. D, Y.Z., and A. I. K. co-wrote the paper. All authors contributed to the results analysis and discussions.

**Competing interests:** The authors declare that they have no competing interests.

**Data and materials availability:** All data needed to evaluate the conclusions in the paper are present in the paper and/or the Supplementary Materials. Additional data related to this paper may be requested from the authors.

**Figures and Tables**

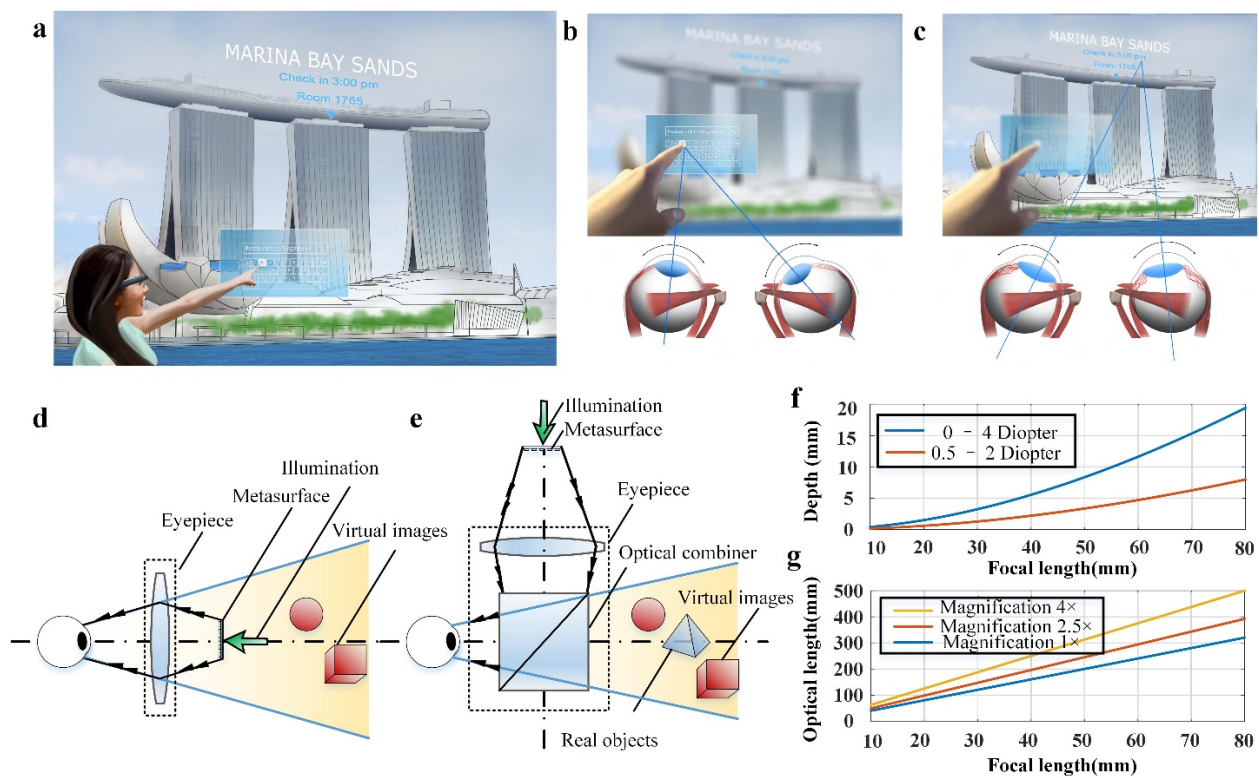

**Fig. 1. General illustration of holographic near-eye displays free from vergence - accommodation conflict. (a)** Schematic of daily life using near-eye displays. A near-eye display will act as a mobile computing device, with virtual information added onto the real world. In the ideal case, virtual and real information will be seamlessly blended in the full, three-dimensional space. **(b)** When the optical axis of both eyes converges to meet the position of near objects, the crystalline lens will also be adjusted to focus on near positions by muscle tension. In this situation, the virtual and real scene in the near area will be clear together, and both virtual and real scene in far area will be simultaneously blurred due to defocusing. **(c)** Same as **(b)**, but when the user is focused on far away objects. **(d) - (e)** Schematic illustration of the general immersive **(d)** and optical see-through **(e)** near-eye display systems using a transmissive metasurface hologram. **(f) - (g)** Relationship between the focal length of the eyepiece and the **(f)** holographic depth and **(g)** overall optical length.

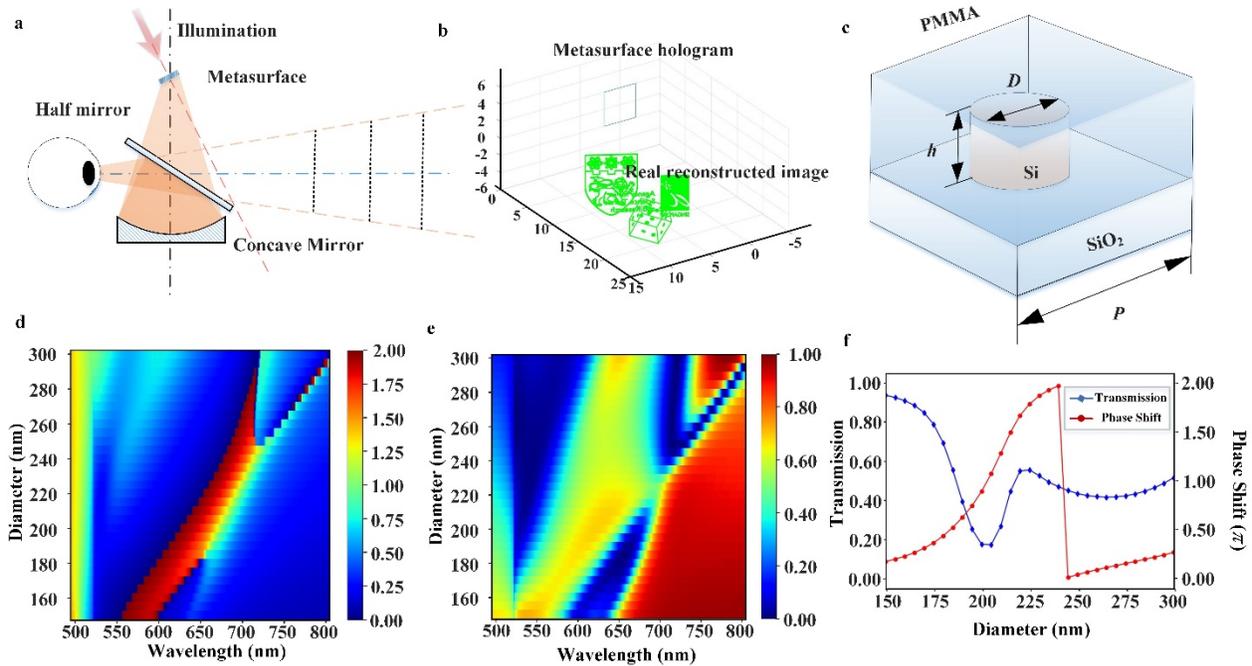

**Fig. 2. System schematic and design of the metasurface hologram. (a)** Illustration of the setup of the optical see-through near-eye display system using a transmissive metasurface hologram. **(b)** Spatial position relationship between the reconstructed 3D images and the hologram metasurface. The function of the eyepiece should be considered to provide correct 3D scene from near area to far distance. **(c)** Schematic illustration of a unit cell of the Huygens' metasurface hologram. **(d)-(f)** Simulated results for regular Huygens' metasurfaces comprising silicon nanodisks of different diameters, with the height of 100 nm and the period of 360 nm embedded into a homogeneous medium with refractive index of 1.5. The colour maps indicate: **(d)** the phase shift of the transmitted wave in π units and **(e)** the transmittance. **(f)** The transmission and phase shift as a function of the nanodisks diameter for a metasurface illuminated by a plane wave at the operational wavelength of 680nm.

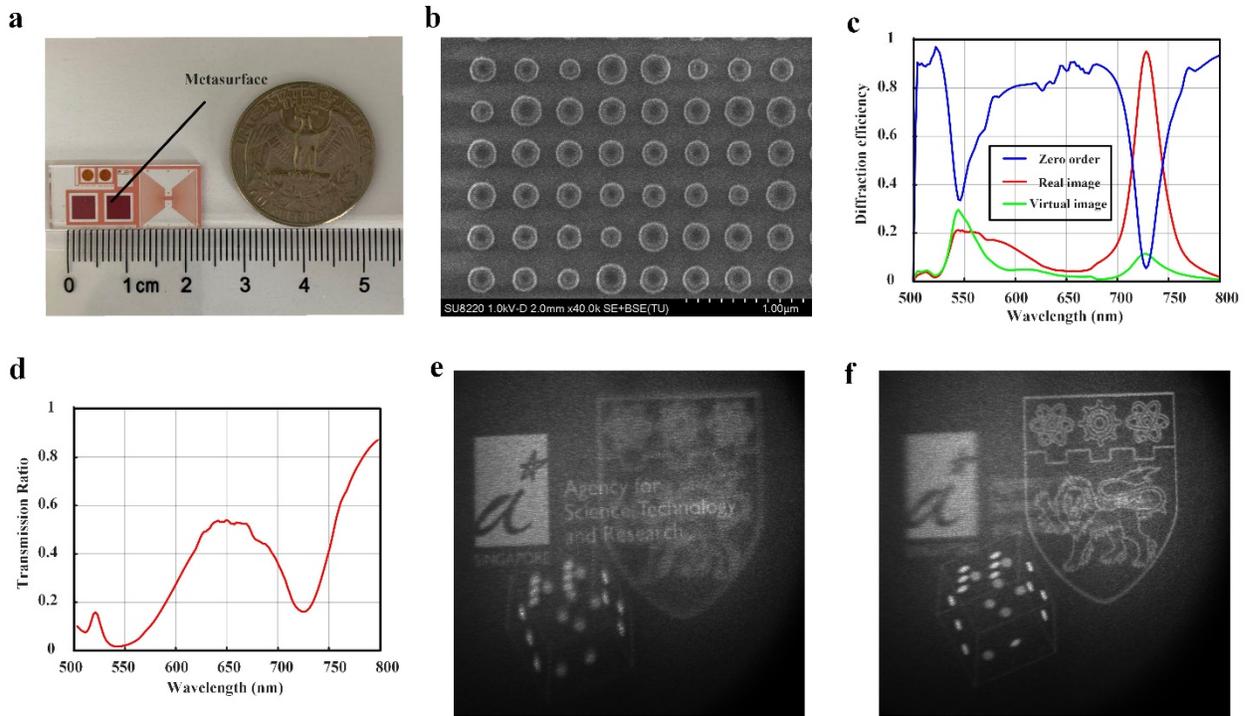

**Fig. 3. Performance of the metasurface hologram. (a)** Photograph of the metasurface hologram sample along with a quarter-dollar coin. **(b)** SEM images of the fabricated sample from top view. **(c)** The diffraction efficiency for zeroth order, real image, and virtual image as a function of the wavelength. **(d)** The total transmission of the metasurface hologram sample as a function of wavelength. **(e) - (f)** Reconstructed virtual 3D images at the wavelength of 728nm captured by the camera when focused at the depths corresponding to **(e)** the A*STAR logo and **(f)** the NTU logo. In **(e)**, the A*STAR logo becomes sharp, while the NTU logo appears blurred due to the defocusing. An opposite situation holds in **(f)**.

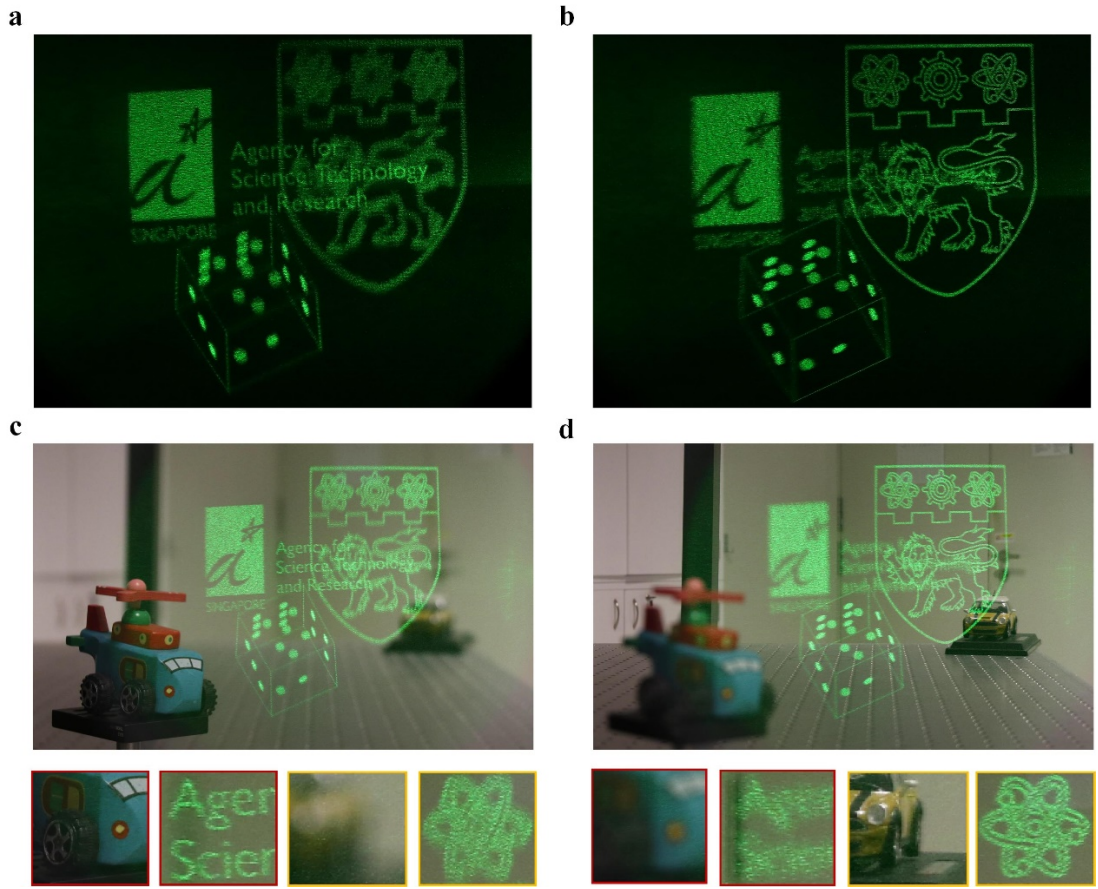

**Fig. 4. Near-eye display using the designed metasurface hologram. (a)-(b)** Virtual 3D images as captured by a visible camera when: **(a)** focused at 2 diopters (corresponding to the distance of 0.5m) and **(b)** focused at 0.5 diopters (corresponding to the distance of 2m). The three virtual objects are reconstructed at different depths ranging from 0.5m (for the A*STAR logo) to 2m (for the NTU logo). **(c)-(d)** Optical see-through scenes as captured by the camera when: **(c)** it is focused at 2 diopters (0.5m) and **(d)** it is focused at 0.5 diopters (2m). In this case, both real and virtual objects coexist in the scene. The holographic image corresponding to the A*STAR logo and the real plane model are at a (same) depth of 0.5m. The virtual NTU logo and the real car model are at a same depth of 2m. The enlarged parts of the AR images are also provided in the bottom raw to clearly show the sharpness and blurriness in details.

**Supplementary Materials**

**Supplementary Note 1.** Design considerations and analysis of a near-eye display set up using a metasurface hologram

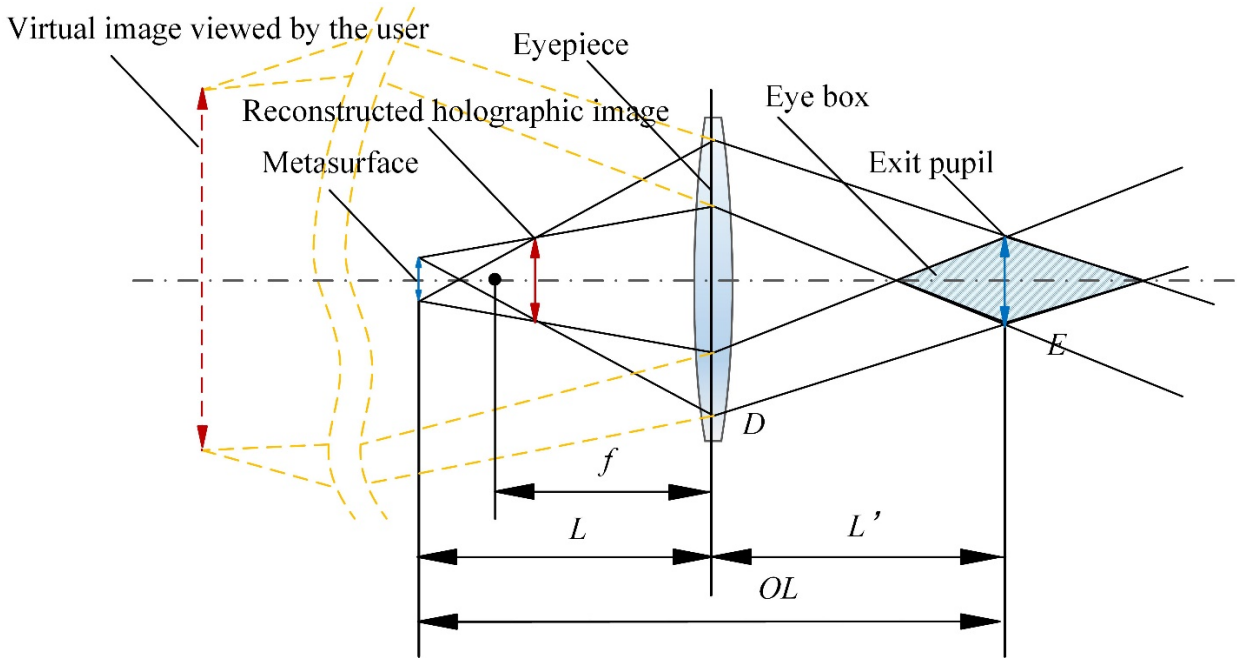

**Fig. S1.** Schematic and specific parameters of a near-eye display using a metasurface hologram.

The metasurface hologram and the exit pupil should be set as object-image conjugate relationship under the function of the eyepiece. The size of exit pupil is $\alpha$ times that of metasurface, and that is to say, the magnification ratio is $\alpha$. The overall length ($OL$) of the system from the metasurface to the exit pupil, the distance ($L$) between the eyepiece and the metasurface, the eye relief ($L'$), the magnification ratio $\alpha$, and the focal length ($f$) of the eyepiece are shown in Fig. S1. The relationship among these parameters is listed as follows:

$$OL = (\alpha+1)^2 f / \alpha$$

$$L' = (\alpha+1) f / \alpha$$

It is assumed that the depth of the 3D scene covers the whole range from near to infinite distances. In the latter case, the light beam is parallel, as is the case for objects viewed from infinite distance. The aperture ($D$) of the eyepiece can be obtained from the field of view ($FOV$), focal length ($f$) and the exit pupil size ($E$) as:

$$D = E + L' \tan(FOV/2)$$

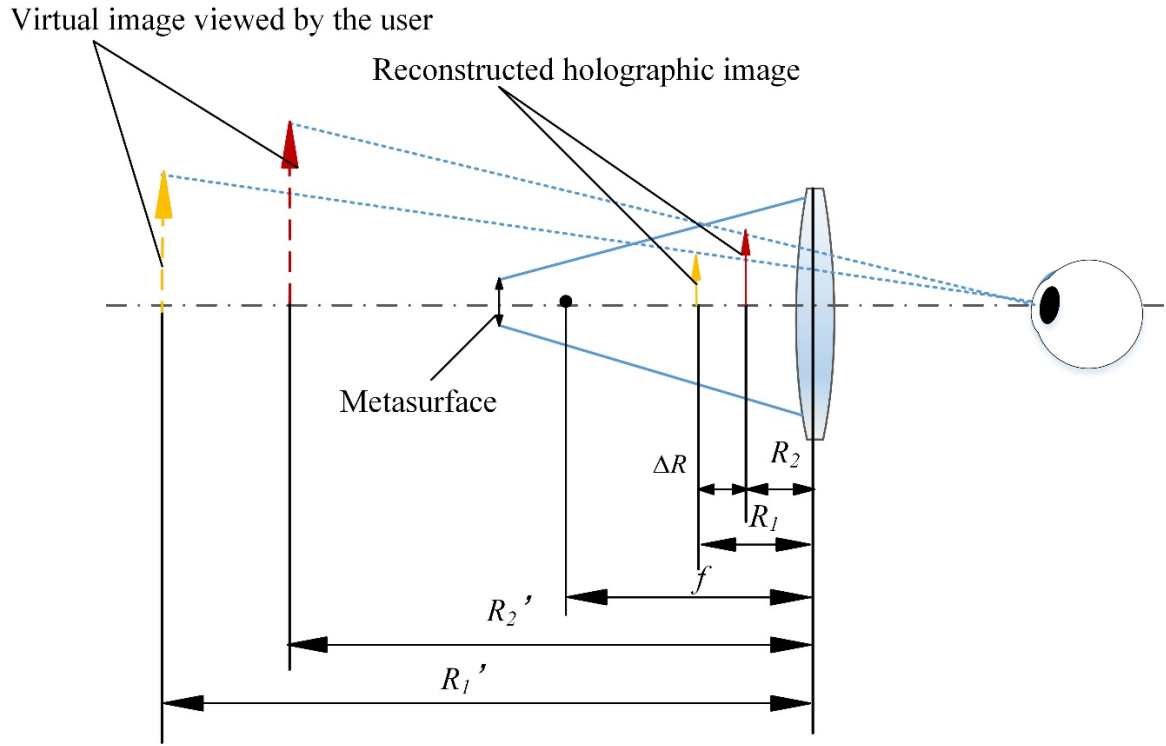

**Fig. S2.** Schematic of the reconstruction depths of the holographic 3D scene provided by the metasurface.

If the required 3D scene covers the depths from position $R_1'$ to position $R_2'$ from the human eye, the distance between the reconstruction depths ($\Delta R = R_1 - R_2$) can be expressed as follows:

$$\Delta R = \frac{(R_1' - L')f}{(R_1' - L') - f} - \frac{(R_2' - L')f}{(R_2' - L') - f}$$

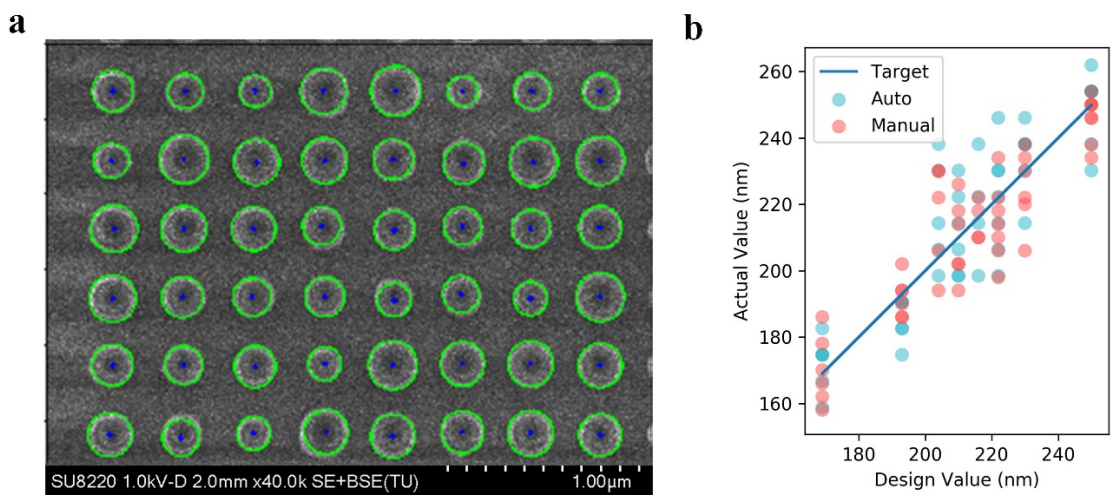

**Fig. S3.** The fabrication error between the target and actual parameters. Intermediate levels can hardly reach the target as the variations are larger than required precision (nm). Two size measurement methods were used, which include manual labeling and automated labeling by computer vision. Comparison between the manual labelling and the automated labelling of sizes does not show large differences. The markers have alpha (transparency) value of 0.3 to reveal overlapping data points.

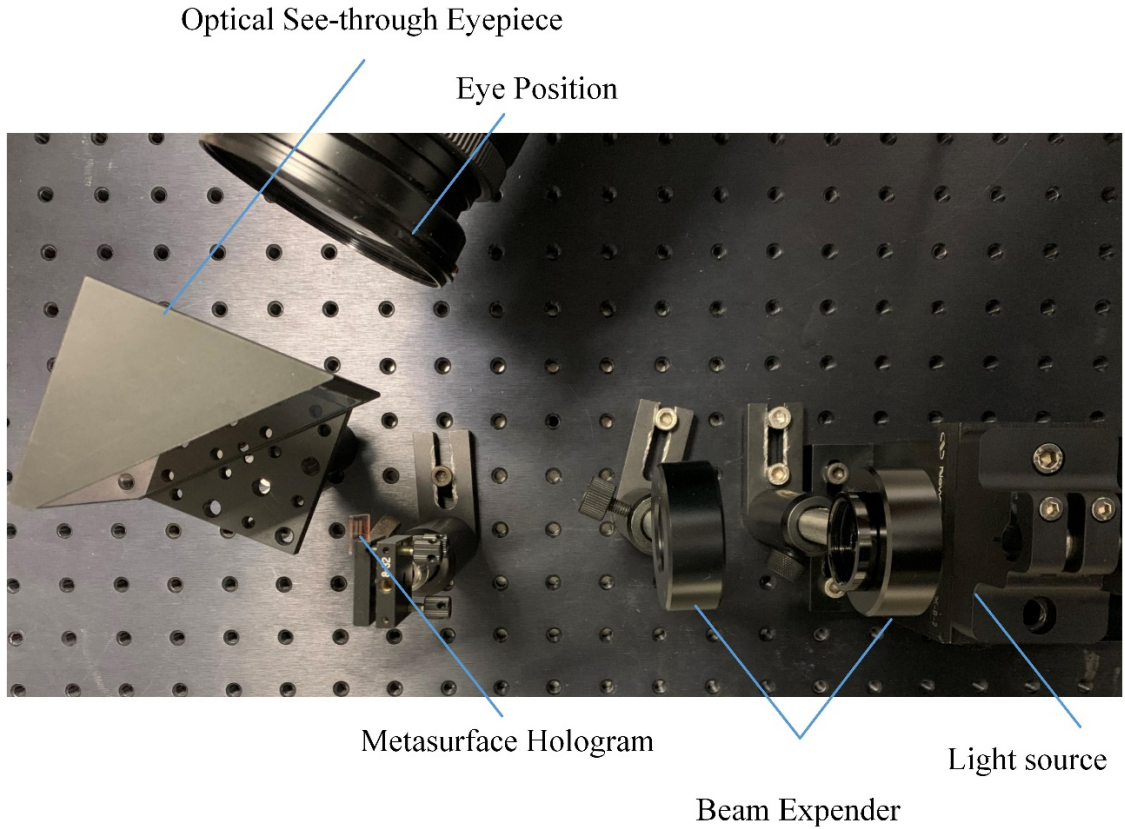

**Fig. S4.** Optical measurement setup. The system consists of a light source (which is a fiber laser), a beam expander (two convex lenses), a metasurface hologram (with a size of 4mm × 4mm), and a custom-made optical see-through eyepiece. The beam expander is used to expand the fiber laser to make the size of the beam suitable for the metasurface hologram. The optical see-through eyepiece is made of one concave mirror and a beam splitter. The virtual 3D scene can be viewed floating in air along with the real world by naked eye within the eye box. To simulate the display performance from the user's eye, a camera is utilized by putting its aperture at the eye's position.

**Supplementary Note 2.** Characterization of the see – through display at different wavelengths

For completeness, the generated images at 530nm, 544nm, 560nm, 580nm and 600nm wavelengths have also been captured by changing the wavelength of the light source and using the camera at the exit pupil. During this process, the positions of all these elements (including the eyepiece, the metasurface, the camera and the source) stayed the same. Only the focus of the camera was adjusted to either focus on the ASTAR logo or the NTU one. The results, as seen in Fig. S5, show that the depth of the image moves away with the increase of wavelength, as expected from a traditional holographic pattern. With the decreasing of absorption and diffraction efficiency, and with the increase of the zero order, the image quality decreases and the speckle noise increases. This tendency continues until reaching again wavelengths near the optimum range at ~728nm.

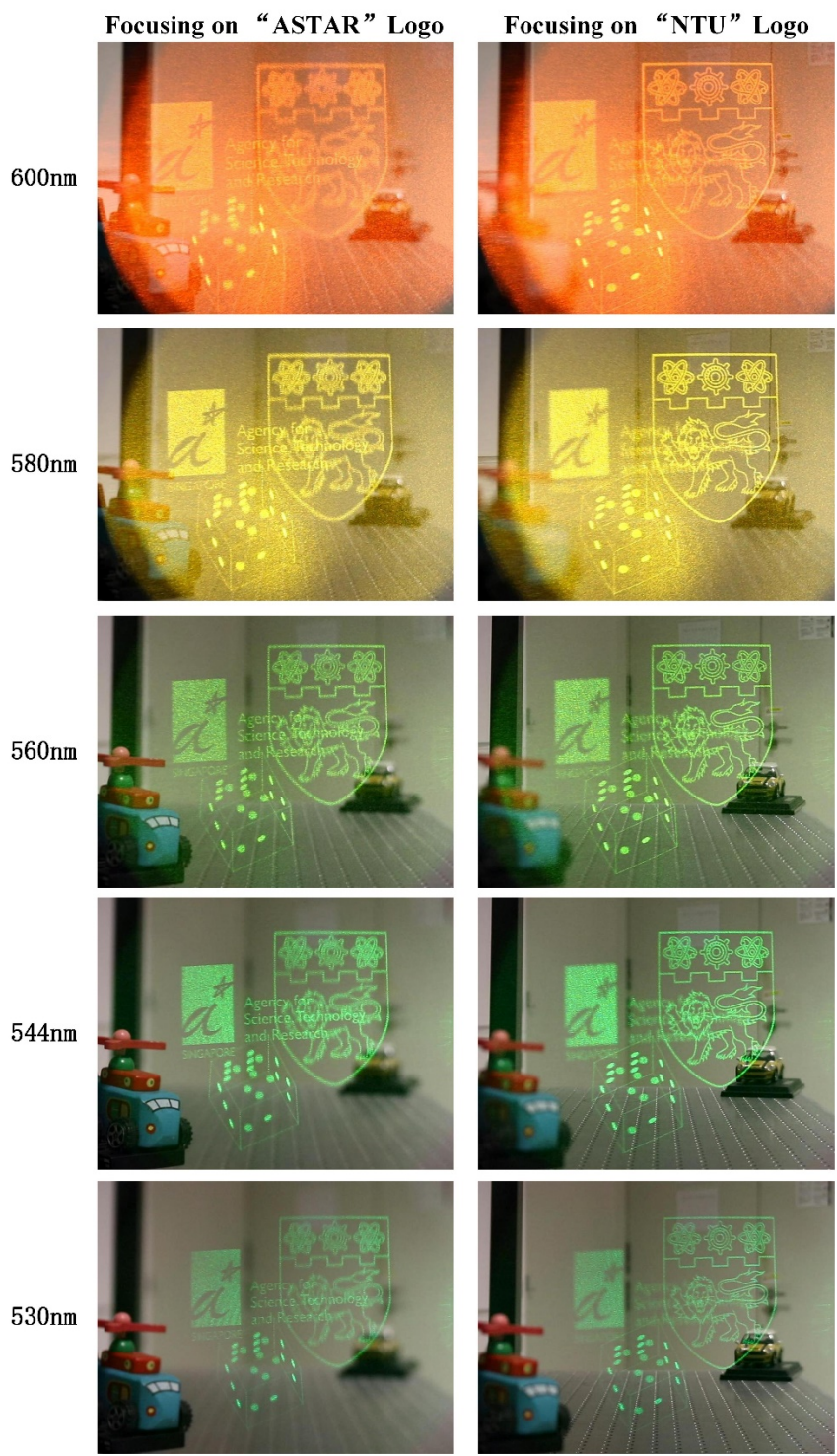

**Fig. S5.** Display performance at different wavelengths. The focus positions are set at the "ASTAR" logo (left column) or "NTU" logo (right column). Holographic images for 530nm, 544nm (the highest holographic efficiency within visible wavelengths), 560nm, 580nm, and 600nm wavelengths are captured. The depth of the image moves away with the increase of the wavelength, which is the same as in traditional holographic patterns. With the decrease of absorption and diffraction efficiency of the metasurface, the image quality deteriorates and the speckle noise increases.